\documentclass[final]{aa}

\usepackage{graphics}

\begin{document}

\thesaurus{11.		      
	   (11.12.1;	      
	   11.09.1 IC\,10;    
             11.09.1 IC\,1613;    
             11.19.5)	      
	   }

\title { Infrared photometry of the
Local Group dwarf irregular galaxy IC 10}

\author{
   J.~Borissova\inst{1}
   \and
   L.~Georgiev\inst{2}
     \and
   M.~Rosado\inst{2}
     \and
   R.~Kurtev\inst{3}
\and
 A.~Bullejos\inst{2}
\and
M.~Valdez-Guti\'errez \inst{4}
}

\institute {Institute of Astronomy, Bulgarian Academy of Sciences 
and Isaac Newton Institute of Chile Bulgarian Branch,
   72~Tsarigradsko chauss\`ee, BG\,--\,1784 Sofia, Bulgaria 
(jura@haemimont.bg)
\and
Instituto de Astronom\'{\i}a, Universidad Nacional Aut\'onoma
   de M\'exico, M\'exico 
 (georgiev@astroscu.unam.mx,\  margarit@astroscu.unam.mx)
\and
Department of Astronomy, Sofia University and Isaac Newton Institute of Chile Bulgarian Branch, James Bourchier Ave. 5, BG\,--\,1164 Sofia,Bulgaria 
(kurtev@phys.uni-sofia.bg)
\and
Instituto Nacional de Astrof\'\i sica, Optica y Electr\'onica, M\'exico
(mago@inaoep.mx)}

\offprints{J.~Borissova}

\date{Received 25 February 1999 / Accepted 16 June 2000 }

\authorrunning {Borissova et al.}
\titlerunning {Infrared photometry of the IC 10.}

\maketitle

\begin{abstract}

We present near infrared $JHK$
photometry for the central area of the irregular dwarf galaxy IC\,10.
Red stellar population of IC\,10 contains  young  (10 -- 50  Myr) massive
red supergiants. Most of AGB stars are younger than 1 Gyr and have
masses up to 12 $M_\odot$. 
Comparing the red supergiants of IC\,10 and IC\,1613
we have determined the reddening of $E(B-V)=1.05\pm0.10$ 
and the dereddened distance modulus $(m-M)_{0}=23.86\pm0.12$ mag of
 Population I stars in IC\,10.
Detection of six  $\rm Br\gamma$  emission structures clearly outline
two star forming regions in the central area of IC\,10.
There is no $\rm H_2$ emissions down to 4.65
$10^{-16}$ $\rm erg\ ñ^{-1} cm^{-2}\  arcsec^{-2}$ indicating that the dominant excitation mechanism of the  molecular gas is from UV radiation
from hot young stars.
From the comparison of  $\rm Br\gamma$ and
$\rm H\alpha$ fluxes, we derive average extinction toward the star forming regions
 $E(B-V)_ {\rm H{\alpha}}= 1.8\pm0.2$. 
The summarized SFR derived from 
six $\rm H\alpha$ and  $\rm Br\gamma$ emission structures in 
our field of view is 0.6  $M_\odot {\rm year^{-1}}$


\end{abstract}

\section{Introduction}

IC\,10 is an irregular dwarf member of the Local group.
This galaxy is an extremely peculiar and interesting object. As pointed
out by Hubble (1936):
"The third nebula, IC\,10 is one of the most curious objects in the
sky. Mayall, at the Lick observatory, was the first to call attention
to its peculiarities...The photographs are difficult to interpret
fully, but they suggest that a portion of a large, late-type spiral
is dimly seen between obscuring clouds".

Shostak et al. (1989) and Wilcots et al. (1998)
have mapped the extended gas around the irregular galaxy IC\,10. They
have shown that the gas around this galaxy is also amazingly
complex. Singe-dish observations of IC\,10 had suggested that the outer
envelope was turbulent and there were velocity gradients at small
and large scales (Cohen 1979), but the new
VLA mosaic shows that the extended HI is concentrated in three arm-like
structures and that IC\,10 is merging with a large infalling
cloud to the south of the disk (Hunter 1997).

The stellar content of IC\,10 was investigated by several authors.
Massey \& Armandroff (1995) found a very well populated blue
plume of main-sequence luminous stars in the $(V,B-V)$ color-magnitude diagram.
They also found  that IC\,10 had a global surface density of Wolf-Rayet
stars 3 times higher than any other
Local Group galaxy. This high concentration of WR stars reinforces the idea
that IC\,10 is undergoing rather strong bursts of star formation.
Saha et al. (1996) using Gunn $g, r$ filters reported the existence
of a blue plume containing the blue supergiants, a red vertical
plume at $(g-r)=1.3$ and other bright red supergiants redward of it.
They also investigated the variable stars in IC\,10 and
reported four Cepheids. Wilson et al. (1996) confirmed the variability
of these stars from IR photometry. Sakai et al. (1998) identify a well
populated red giant branch, an ill-defined blue main sequence population
and intermediate-age asymptotic giant branch. In latest works of 
Hodge \& Zucker (1999), Hunter (1999) and Tikhonov (1999) stellar content,
reddening and distance are also analyzed.
Each of the above works however reported different values of the distance modulus and
the reddening, varying from $E(B-V)=0.75$ to $1.16$ and $(m-M) = 22.1$ to
$24.9$.
The calculations of reddening and distance reported above are based on
different stellar population types ---
Wolf-Rayet stars (Massey \& Armandroff 1995),
Cepheid variables (Saha et al. 1996; Wilson et al. 1996) and the tip of
the red giant branch method ( Sakai et al. 1998,  Tikhonov 1999).
The most probable explanation of these so different results is that IC\,10
lies extremely close to the Galactic plane thus having
a high optical foreground reddening.
Another possible reason for these differences could be the fact that
since this is a starburst galaxy it has a variable internal reddening.

This paper presents $JHK$ 
photometry of the central area of IC\,10.
The purposes of our work are to determine the stellar content, the reddening,
the distance modulus and the ages of different stellar populations in IC\,10
by means of an approach which is less
sensitive to both the foreground and the internal reddening
and is at the same time sensitive to the star formation activity
--  near IR observations.

Our observational material and data reduction 
techniques are described in Section 2 together with
 our analysis of the photometric errors and completeness
 corrections.  Several aspects of the color-magnitude diagrams 
(CMD) of IC\,10 including analysis of stellar content are described
 in Section 4. IC\,10's reddening, distance modulus and 
age are determined in Section 5 and  Section 6.  
The recent star formation activity based on the $\rm Br\gamma$ 
emission is analyzed in Section 8 and in Section 9 
we summarize the most important results of the present investigation.

\section{Observations and data reduction}

\subsection{$JHK$ and $UBV$ observations}

The data discussed here were acquired with the
Infrared Camera "CAMILA" (Cruz-Gon\'alez et al. 1996)
with a NICMOS3 ($256\times 256$ pixels)
detector  attached to the
2.1-m telescope of the Observatorio Astronom\'{\i}co
National "San Pedro Martir", M\'exico. The scale was $0.85$
arcsec/pixel, resulting in a field size of about $3.6$ by $3.6$ arcmin.
The total exposure times are 600 sec in $J$, 480 sec. in $H$ and 640 sec.
in $K$.
A set of $JHK$  frames was taken on January 11-13, 1998.
$JHK$ images of the Local group irregular galaxy IC\,1613 were
being taken for comparison purposes on the same nights. 
An additional set of 
$BV$ frames of IC\,10 and Off-field (a field 30 arcmin north)
 were obtained on the 2-m Ritchey-Chretien
telescope of the Bulgarian National Astronomical Observatory
on September 15, 1999 with a Photometrics $1024\times 1024$ camera. 
The exposure times of optical images are 900 sec. 
The scale at the Cassegrain-focus CCD was 
$0.33\arcsec\,{\rm pixel}^{-1}$ and the observing area 
was $5.6\arcmin \times\ 5.6\arcmin$ centered on the same field as in infrared images.
The  seeing during these observations 
was between $1 - 1.2 \arcsec$ with stable and very good photometric 
conditions. Twelve UKIRT (Casali \& Hawarden  1992)
 standard stars as well as Landolt (1992) standards
were taken before and after all observations.

\begin{figure}[htbp]
\resizebox{\hsize}{!}{\includegraphics{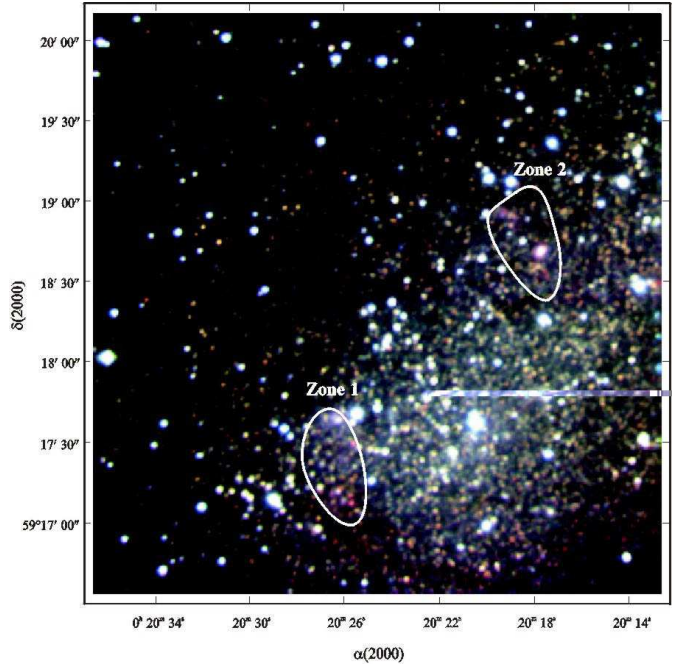}}
 \caption{ The red-green-blue composite true color image generated from the
 J (blue), H (green) and K (red) images. Two regions with
infrared excess are marked as Zone 1 and Zone 2.
}
 \label{Fig01}
\end{figure}

A red-green-blue composite true color image generated from  J (blue), H
(green) and K (red) images is shown in Fig.~\ref{Fig01}.
%

The IRAF data reduction package was used to carry out the basic image
reduction. The stellar photometry of the frames was done using DAOPHOT 
(Stetson 1993).

\begin{figure}[htbp]
	    \resizebox{\hsize}{!}{\includegraphics{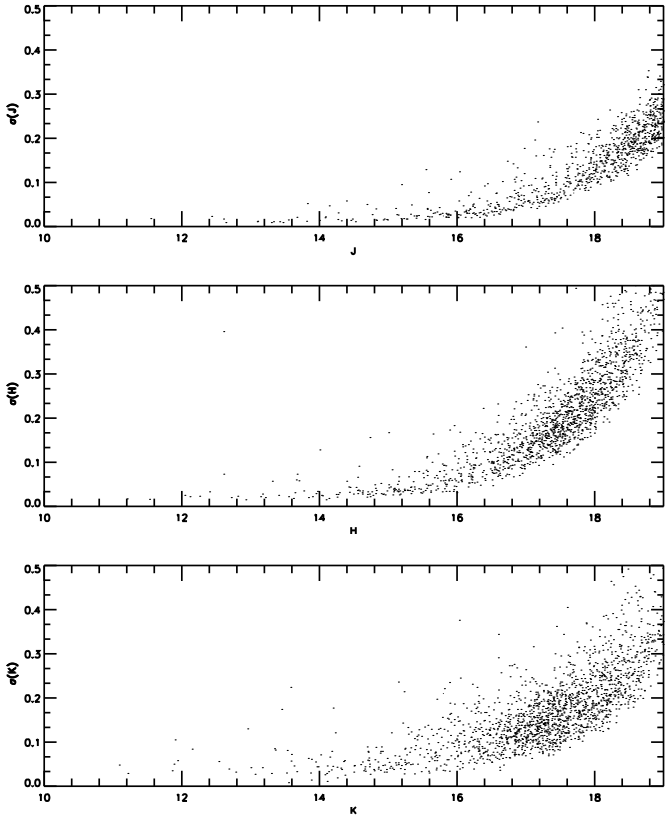}}
      \caption{Internal errors of the IC\,10 photometry.}
      \label{Fig02}
\end{figure}

     In order to estimate the internal accuracy of our photometry we used
the formal errors from the DAOPHOT package.
For all stars the standard errors vs. magnitudes are displayed in Fig.~\ref{Fig02}.
As can be seen we do not have large errors $(> 0.25)$ up to the 
magnitudes: $ J = 18.0$, $ H = 17.5$ and $ K = 18.5$. Only the stars brighter than these limits are used on the color-magnitude diagrams.

\begin{figure}[htbp]
   \resizebox{\hsize}{!}{\includegraphics{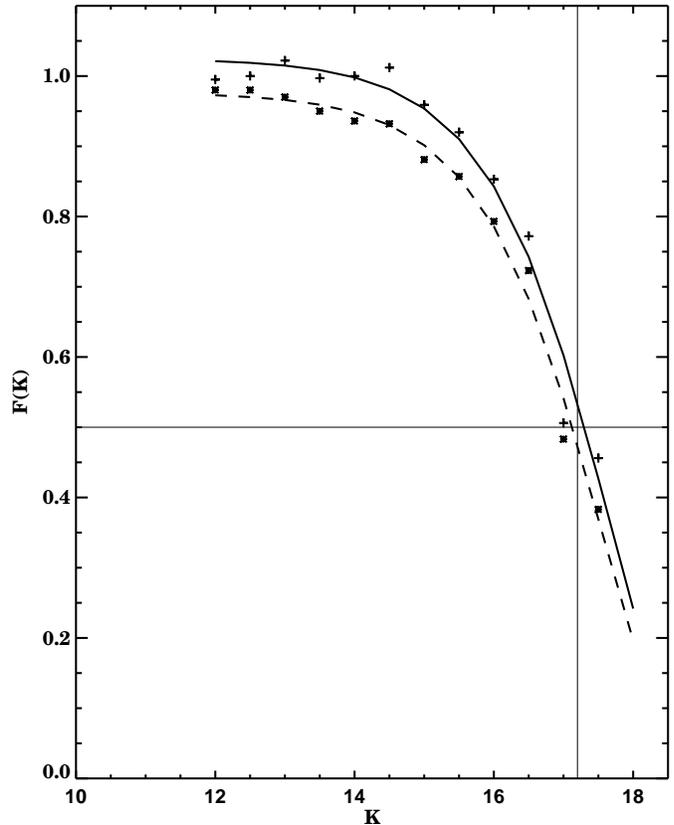}}
\caption{
 Completeness curves in annuli at different distances from the galaxy center
as a function of the $K$ magnitudes. Solid line represent completeness 
function for outer region ($ r >  1.80 $
arcmin) and the dashed one is for  ($r < 1.80$ arcmin).}
\label{Fig03}
\end{figure}

The artificial star technique (Stetson 1991) was used to outline the limits
of our photometry and series of artificial frames were reduced in the same
manner as the original frames. 
Since completeness is a function of the distance from the galaxy
center and of the limiting magnitude, we performed the procedure in two
radial annuli with  $ (r <  1.80) $ arcmin and
$ (r > 1.80) $ arcmin.  Fig.~\ref{Fig03} shows the completeness functions in the two
different annuli plotted as a function of the K magnitude.
As can be seen, $50\%$ completeness limit is at $K \approx 17.2$ for the 
both zones. 
The analysis of Fig.~\ref{Fig03} also shows that
the large pixel of the IR detector can affect the detection of the bright stars.
The number of detected stars for the fixed magnitude in the outer region is  $\approx 5\%$
larger than in the inner region.
We estimate our $50\%$ completeness limit to be 17.0 magnitudes in $J$ and 
$H$ filters.

%

\subsection{$\rm Br\gamma$ and $\rm H_2$ observations}

Narrow band imaging in the $2.17\mu \rm m$ $\rm Br\gamma$,
$2.12\mu \rm m$ $v=1-0$ S(1) $\rm H_2$ and $2.26\mu \rm m$
continuum lines was carried out on the last night of the "San Pedro Martir" 
observing run  --- 13 January 1998.
The atmospheric conditions were stable. Flux calibration was made using the
WD atmosphere model of the faint ESO DA0 spectrophotometric standard
star G191-B2B (Oke 1990, Bohlin et al. 1995). Images of the standard
star were taken at the same airmasses as the program fields.

\section{Comparison with previous works }

Unfortunately there are only two papers on the IR-photometry of IC\,10 and
IC\,1613 and they deal with a few stars only  --- Wilson et al. (1996)
and Elias \& Flogel (1985).

There is only one star from our field of observations that is
present in the infrared IC\,10 data of Wilson et al. (1996) --- the very
red variable star V3.
The difference in $J-K$ color is $(J-K)_{\rm our}-(J-K)_{\rm Wilson} = 0.01$
and in $K$ magnitude is $K_{\rm our}-K_{\rm Wilson} = 0.06$.

There are two stars common to our and Elias \& Flogel's (1985) IR photometry
of IC\,1613 --- the well known variable red supergiants V32 and V38
(Sandage 1971).
Elias \& Flogel (1985) give observed colors and magnitudes for V32 and V38
$J-K = 0.87, K=13.12$ and $J-K = 0.84 , K = 13.12$, respectively.
The differences
between our data and that of Elias \& Flogel are  0.1 in $K$ and 0.05 in $J-K$
for both stars.

Taking into account that V3 in IC\,10 and V32 and V38 in IC\,1613 are variable
stars we have a good agreement between our data and that of Wilson et al. (1996)
and Elias \& Flogel (1985).

\section{ Color-magnitude diagrams}

\begin{figure}[htbp]
	   \resizebox{\hsize}{!}{\includegraphics{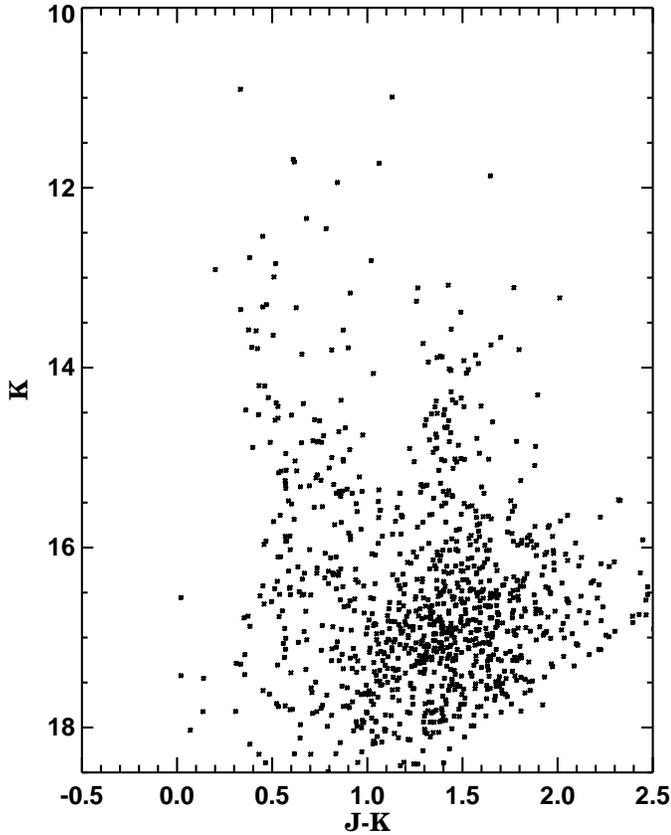}}
      \caption{ $(J-K,K)$ color-magnitude diagram obtained for IC\,10.}
	   \label{Fig04}
\end{figure}

\begin{figure}[htbp]
	   \resizebox{\hsize}{!}{\includegraphics{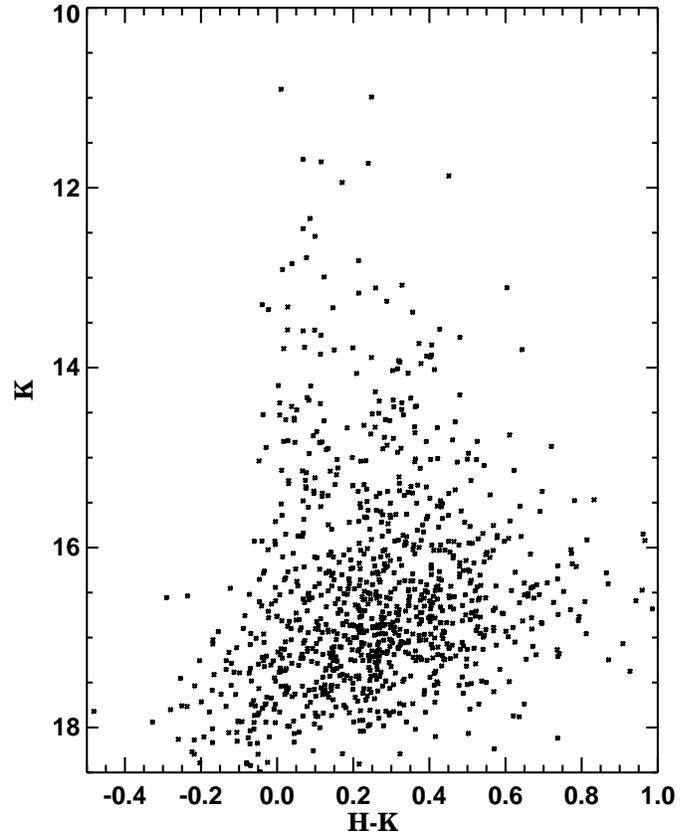}}
      \caption{ $(H-K,K)$ color-magnitude diagram obtained for IC\,10.}
      \label{Fig05}
\end{figure}

      Fig.~\ref{Fig04} and Fig.~\ref{Fig05}  show the 
$(J-K,K)$ and  $(H-K,K)$ color-magnitude
diagrams obtained for IC\,10. Only stars with photometric
errors less than $0.25$ were selected in all filters. Our final list contains
1036 sources with $J$, $H$ and $K$ magnitudes. Data are available 
by e-mail on request to the first author.

%

%

\subsection{ Field star contamination}

IC\,10 is situated close to the galactic plane ($b=-3\fdg 34,\ l=118\fdg  97$) and the foreground contamination is significant.
It is obviously necessary to check the level of field star
contamination before discussing the detailed structure of the color-magnitude 
diagrams and the stellar content of the IC\,10.

\begin{figure}[htbp]
	   \resizebox{\hsize}{!}{\includegraphics{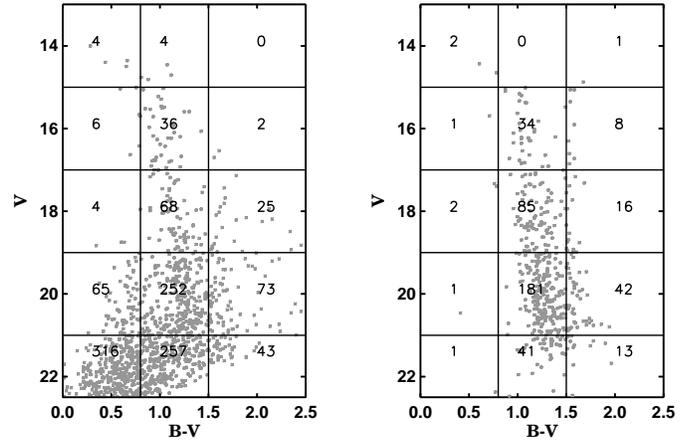}}
      \caption{ $(B-V,V)$ color-magnitude diagram for IC\,10 (left panel).
The Off-field images (right panel) were taken 30 arcmin north of the center 
of galaxy with the same exposure time (900 sec in both filters).
The grid of 15 boxes and number of stars counted in each of them are shown.
  }
      \label{Fig06}
\end{figure}

The ($B-V, V$) color-magnitude diagrams of IC\,10 and an Off-field image are shown in
Fig.~\ref{Fig06}. The  Off-field images were taken 30 arcmin north of the center of the galaxy with the same exposure time as the galaxy field (900 sec in both filters).
As can be easily seen most of the stars with  $1.5 >B-V > 0.8$
are field stars. The tip of the IC\,10 main sequence stars is located close
to $B-V=0.6$ and $V = 18.5$. The few most luminous stars from the red
plume are located between $21 > V > 18$ and $B-V > 1.5$.
These results agree with the analysis of Hodge \& Zucker (1999) who obtained
$UBV$ photometry with 3.5 m. Apache Pont Telescope.

%

\begin{figure}[htbp]
	   \resizebox{\hsize}{!}{\includegraphics{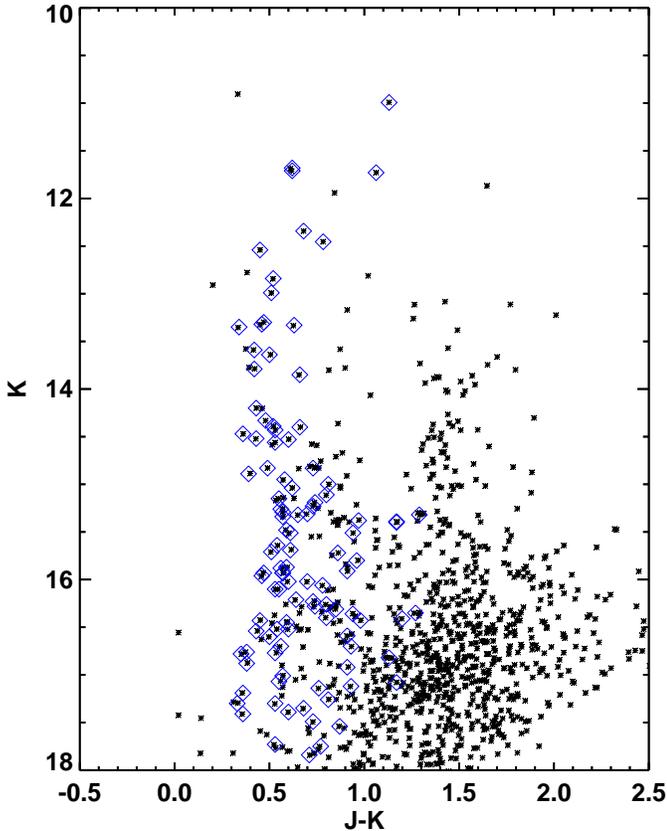}}
      \caption{ $(J-K,K)$ color-magnitude diagram for IC\,10.
The selected field stars are shown as open diamonds.
}
      \label{Fig07}
\end{figure}

To decontaminate the $(B-V,V)$ color-magnitude diagram of IC\,10 we divided
the two CMDs, "galaxy+field" and "field"  into 15 boxes (shown in
Fig.~\ref{Fig06} ) - five intervals in magnitude and three in color. 
The stars in each box in the two diagrams are counted. Then, an
equivalent number of stars is removed from the single boxes of the diagram
of "galaxy+field", on the basis of the number of field stars found in the
diagram of the "field" alone.  Such obtained list of statistically
selected field stars in the  $(B-V, V)$ color-magnitude diagram   
were cross-identified with the IR photometry data
and were superimposed on the $(J-K,K)$ color-magnitude diagram
in Fig.~\ref{Fig07} as open diamonds.

%

The analysis of Fig.~\ref{Fig07} shows that most of the stars
between $0.4<J-K<1.0$ and $H-K<0.1$ (see Fig.~\ref{Fig04} and Fig.~\ref{Fig05})
are field stars while stars with $J-K$ colors greater than 1 and $H-K>0.1$
belong to IC\,10 and only few field stars fall in the latter zone.
The red supergiants  stand around $J-K=1.4$ and $13.0<K<15.0$ mag.  
The AGB stars are visible at $J-K>1.1,\ 0.2<H-K<0.4,\ K >15.0$. 
The red giant population can be seen at the
limit of our photometry at $ K=17.5$.
Our photometry is not deep enough to see any IC\,10 
main sequence stars there.

The observed $J-K,K$ and $H-K,K$ color magnitude diagrams are decontaminated by  subtracting
the selected field stars (open diamonds in Fig.~\ref{Fig07}) from the list of photometry. The remaining $\approx 20\%$ of the stars between $J-K<1.0$, $H-K<0.1$ and $K<16.5$ are considered to be the field stars and are also removed.

\section {Reddening and Distance}

\begin{figure}[htbp]
	   \resizebox{\hsize}{!}{\includegraphics{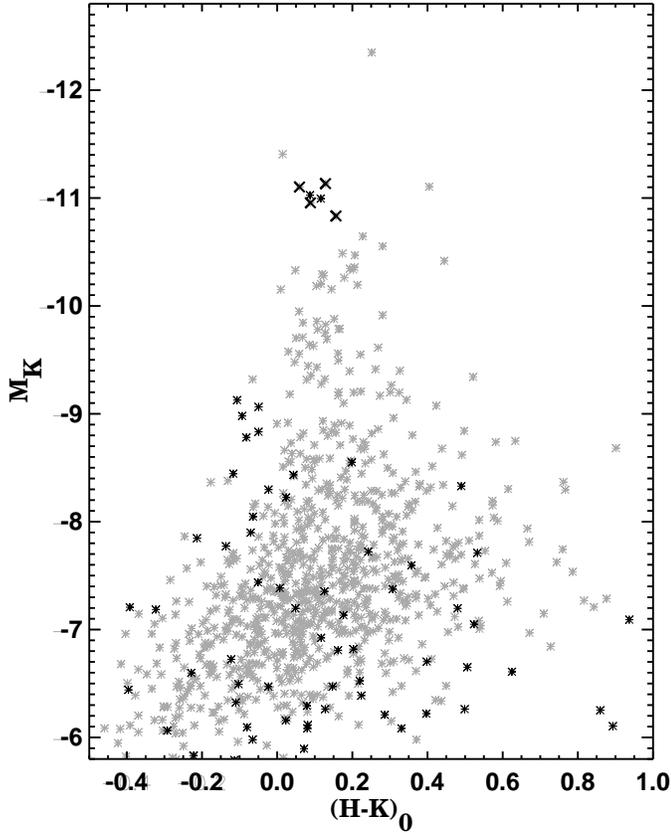}}
\caption {Superimposed color - magnitude diagrams of IC\,10 (light asterisks)
and IC\,1613 (dark asterisks). The four brightest red supergiants used for comparison (see the text)
are marked by crosses.
IC\,1613 is corrected with $E(B-V) = 0.06$ and $(m-M)_{0}=24.2)$.
IC\,10 is corrected with $E(B-V)=1.05$ and $(m-M)_{0}=23.86$. 
Field stars of IC\,10 are omitted from the plot.}
\label{Fig8}
\end{figure}

A common method of determination of reddening and distance is
the comparison with another well known galaxy. In our case we used
as a comparison galaxy IC\,1613. 
The selected area of IC\,1613 is centered on the HII regions of the
northeast sector. 
IC\,1613 has a very low reddening $E(B-V)=0.03 - 0.06$
(Sandage 1971; Freedman 1988; Georgiev et al. 1999),  the field
contamination from our Galaxy is negligible (Georgiev et al. 1999)
and it has a relatively well known distance ---
$(m-M)_0 = 24.2\pm0.2$ (Freedman 1988, Saha  et al.
1992).
We choose to compare red supergiants of two galaxies. 
Elias et al. (1981) demonstrated that infrared photometry of
M supergiants does not provide by itself reliable distances
to galaxies because the stellar $K$ magnitude depends on galaxian
luminosity and metallicity. In our case we have two galaxies with
similar luminosity and metallicity. According to Mateo (1998) IC\,10 has a
total magnitude $M_V=-15.7$ and $12+{\rm log(O/H)} =8.19\pm0.15$ and IC\,1613 has 
$M_V=-14.7$ and $ 12+{\rm log(O/H)} =7.80\pm0.20$. Most of
the red supergiants are known variables at visual wavelengths but there are
no large variations in the infrared. Typical amplitudes are about 0.1 mag or
less in $K$.
Irrespective
of the very similar metallicities derived for both galaxies the ($K,H-K$)
color magnitude diagrams were used to minimize the
metallicity effects.
The mean $H-K$ and $K$ values for the two red variables V38 and V32 of IC\,1613
were corrected for reddening ($E(B-V)=0.06$) and compared with the mean
values of the brightest four red supergiants in IC\,10 standing between $0.2<H-K<0.4$
and $13.0<K<13.7$ (marked in  Fig.~\ref{Fig8} as crosses).
Based on the assumption that reddening is the only reason for the differences
between the colors of the red supergiants of the two galaxies we derived
$E(H-K) = 0.20\pm0.06$ mag.
This value corresponds to $E(B-V)=1.05\pm0.10$ mag,
(the error is our conservative estimate) using the relation
$E(H-K)/E(B-V) = 0.19$ (Bessell \& Brett 1988).
The calculated distance modulus $(m-M)_{0} = 23.86\,\pm0.12$ for
IC\,10 was obtained from the corresponding differences in 
the dereddened $K$ magnitudes. These
values are very close to the values reported by Sakai et al. (1998) for Population I stars
($E(B-V)=1.16\pm0.19$, $(m-M)_{0} = 24.10\pm0.19)$.

The superimposition of the color-magnitude diagrams of the
two galaxies is shown in Fig.~\ref{Fig8}.

%

\section {Age}

The next step in our analysis was to determine the age interval of IC\,10.
We used the ($J-K,K$) color magnitude diagram assuming $z=0.004$, $E(B-V)=1.05$ 
and $(m-M)_{0} = 23.86$
and superimposing the isochrones for $10^7$, $2.5\,10^7$, $2.5\,10^8$ and
$10^9$ years from Padua's library (see Bertelli et al.
1994).

\begin{figure}[htbp]
	   \resizebox{\hsize}{!}{\includegraphics{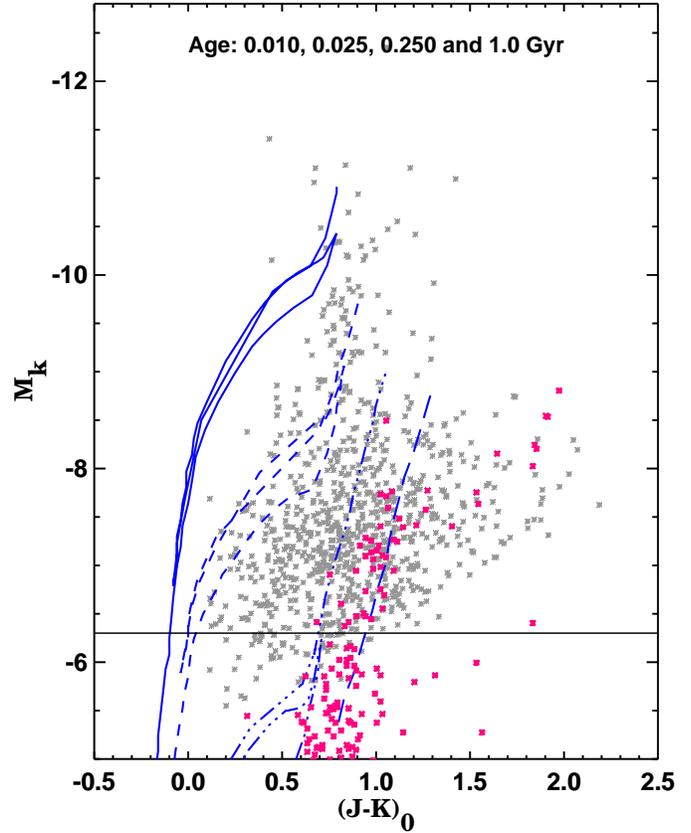}}
 \caption{ The ($M_K,(J-K)_{0}$) color-magnitude
diagram of IC\,10 with superimposed isochrones from Padua's library. The
dark  asterisks show the photometry of the stars  in 12  Large Magellanic Clouds  globular clusters (Ferraro et al. 1995), corrected with $E(B-V)=0.10$ and $(m-M) = 18.60$. The horizontal line locate the separation threshold between AGB (upper part) and RGB (lower part) stars for these clusters as given in Ferraro at al. (1995). Field stars of IC\,10 are omitted from the plot.
}
 \label{Fig9}
\end{figure}

The analysis of the superimposed isochrones (Fig.~\ref{Fig9}) indicates that
the red supergiants have ages between 10 and 50 Myr. The unambiguous
presence of young red
supergiants population corresponds to the presence of a large number of
WR stars and the hypothesis of recent star burst formation. 
Most of AGB stars have ages less than 1 Gyr and can be identify at $(J-K)_{0}=0.6$. There is a significant presence of AGB stars around 1 Gyr also. Some older AGB stars with ages of several Gyr can be seen. 

In the same plot $JK$ photometry of the stars in twelve Large Magellanic Clouds globular clusters  (Ferraro et al. 1995) are also given (dark  asterisks in  Fig.~\ref{Fig9}). The horizontal line locate the separation threshold between AGB (upper part) and RGB (lower part) stars as given in Ferraro at al. (1995). The comparison with globular clusters AGB stars confirms the presence of early AGB stars in IC\,10.

%

\section {Theoretical H--R diagrams}

\begin{figure}[htbp]
	    \resizebox{8cm}{!}{\includegraphics{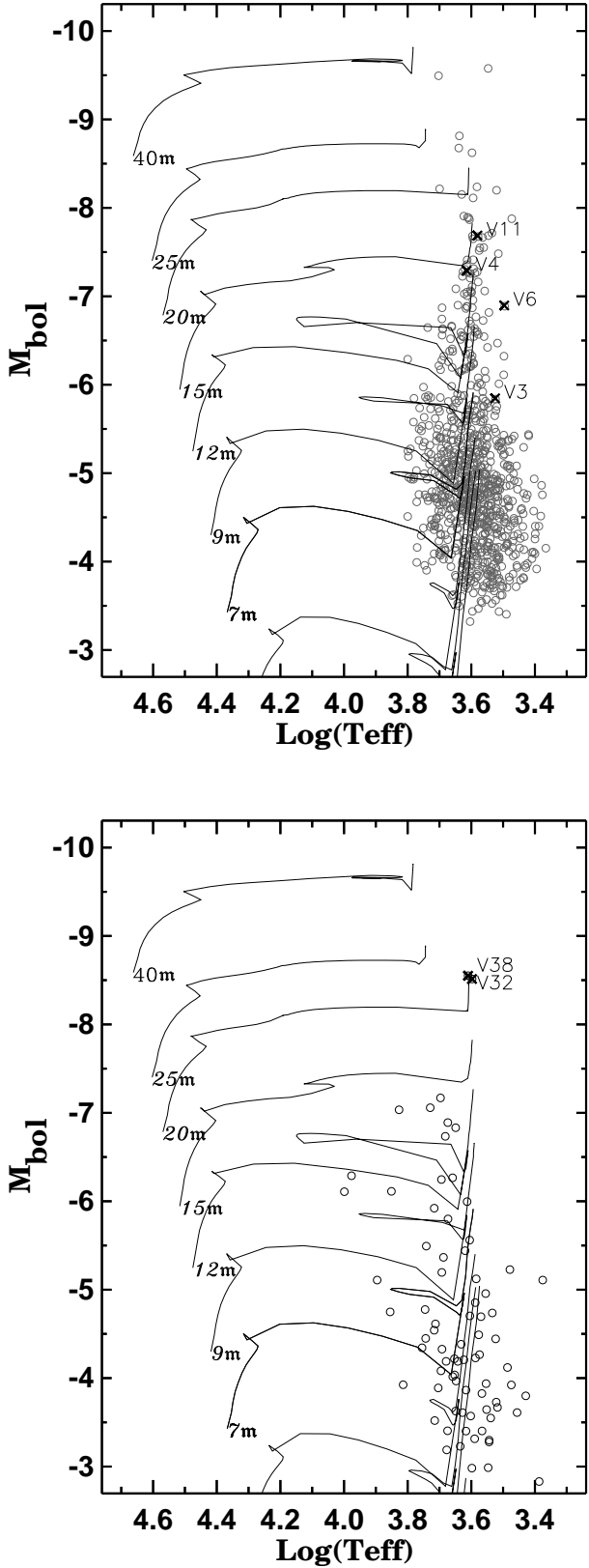}}
      \caption{ H--R diagram for all stars in IC\,10 (upper panel) and IC\,1613 (lower panel).
 The evolutionary mass tracks from Charbonnel et al. (1993) for $z=0.004$ 
are superimposed on the same plots. The variable stars are marked by crosses. Field stars of IC\,10 are omitted from the plot.}
      \label{Fig10}
\end{figure}

One of the best ways to analyze the stellar content of a galaxy
is to put its stars on the theoretical H--R diagram.
The comparison with theoretical evolutionary mass tracks predicted by
theoretical models will give us the evolutionary stage of each star.
To plot the stars on the theoretical H--R diagram we need to determine their
effective temperatures $T_{\rm eff}$ and bolometric corrections (BC).
We used  Costa \& Frogel's (1996) method
to transform K magnitudes and $J-K$ colors into $M_{\rm bol}$ and
$T_{\rm eff}$.
We first transformed our $K$ magnitudes and  $J-K$ colors into the
CIT system by means of equations  1 and 2  derived by Ruelas-Mayorga (1997)
for the Mexican set of filters.
Then using equation 6 from Costa \& Frogel (1996) we transformed
our $(J-K)_{CIT}$ colors to the Johnson  system. The bolometric correction
for the K magnitude was calculated using their equation 1,
while $T_{\rm eff}$ for
each star was derived as the mean value of the temperatures given by their
equations 8 and 9.
The evolutionary mass tracks from Charbonnel et al. (1993) for
$z = 0.004$ were superimposed on the same plot. For IC\,10 we used
the distance modulus and reddening derived in Section 5.
For IC\,1613 -- our comparison galaxy,
the reddening and distance are  $E(B-V)=0.03 - 0.06$
(Sandage 1971; Freedman 1988; Georgiev et al. 1999) and
$(m-M)_0 = 24.2\pm0.2$ (Freedman 1988, Saha  et al.
1992) respectively.
The resulting H--R diagrams  are shown in Fig.~\ref{Fig10}.

%

Analysis of H--R diagram of IC\,1613 show that the two 
red irregular variables V32 and V38 (Sandage 1971) are the brightest representative
of the red supergiants in this most active area of the galaxy. The derived effective temperatures 
of V32 and V38 are approx. 4000 K, which is in agreement with their M0Ia 
spectral class (Elias \& Flogel 1985). Some early AGB  or/and RSG  can be identify at 
$-6.5 <M_{\rm bol} < -7.5 $ and 12 to 15 $M_\odot$. 
The TRGB can be found around $M_{\rm bol} = -4.0 $. Deeper photometry is necessary to 
investigate the red giant branch in details.

To analyze the variable stars in IC\,10 we used the 
Saha et al. (1996) and Wilson et al. (1996) investigations of the light curves
and periods for 13 variable stars. Four of them were found to be Cepheids.
Unfortunately only four of these variables are measured in our photometry
--- V3, V4, V6 and V11. There are no Cepheid variables among them. V4 and V6
were found by Saha et al. (1996) to be eclipsing variables. Sakai et al (1999) reported
V6 as a shortest-period Cepheid with $P=8$ days, but with very anomalous color.
In our H--R diagram this star also have lower than normal Cepheid temperature.
V3 is a very red variable
star with $P=7.1123$ whose status is unclear. V11 stands in the
$(r-i,r)$ diagram of Saha et al. (1996) (see their Figure 7) among the red
supergiants and has been classified by Saha et al. (1996) as a Cepheid-like
variable with a very long period ($P = 90.70$ days).
In Fig.~\ref{Fig10} these stars are marked by crosses. 
As can be seen V11 has
$M_{\rm bol} = -7.69 $, $\rm log(T_{\rm eff}) = 3.58 $ and 15 $M_\odot$.

Taking into account the analysis of IC\,1613 and position of the variable stars on the H--R diagram of IC\,10 we consider the stars with $M_{\rm bol} < -7 $,
$ 3.5 < {\rm log(T_{eff}}) < 3.7 $ and masses greater than 15 $M_\odot$
to be red supergiants. Some early AGB stars and/or RSG can be identify at evolutionary 
tracks with 12 $M_\odot$.  Intermediate age AGB stars stand 
between 7 to 9 $M_\odot$. The TRGB can be found at $M_{\rm bol} = -3.8 $. As in IC\,1613 
 deeper photometry is necessary to investigate the red giant branch in details. 

\section {Survey of star forming regions in IC\,10}

The near infrared $JHK$ images can be used as continuum images which represent
stellar light and hot dust emission from HII region complexes and dust
extinction.
On the red-green-blue composite true color image generated from J (blue), H
(green) and K (red) images (Fig.~\ref{Fig01})  several structures with
infrared excess can be identify. They can be divided in two regions
named Zone 1 and Zone 2.
The reddest object in Zone 1 is rather compact and has a mean diameter of
6 arcsec. Zone 1 also contains two fainter, extended and diffuse structures.
Several stellar-like objects embedded into diffuse structures can be identify
in Zone 2. 

\begin{figure}[htbp]
	   \resizebox{\hsize}{!}{\includegraphics{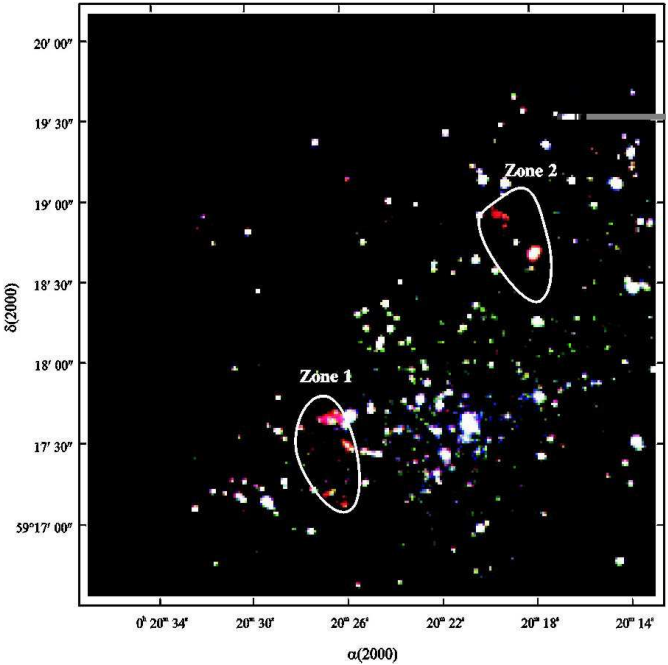}}
 \caption{ The red-green-blue  composite true color image generated from the
  $\rm H_2$ --- $\rm 2.122\mu m$ (blue), continuum --- $2.26\mu m$ (green) and
 $\rm Br\gamma$ --- $2.165\mu m$ (red) images. North is at the top and east to the  left. 
The field of view as shown is 3.6 arcmin. Two regions with the $\rm Br\gamma$
emission structures are marked as Zone 1 and Zone 2.
}
 \label{Fig11}
\end{figure}

It is well known that the $\rm Br\gamma$ recombination line of hydrogen
traces sources of Lyman continuum flux i.e. hot young stars, while $\rm H_2$ 
arises from hot molecular gas and traces the material available for star formation.
A red-green-blue  composite true color image generated from the
$\rm 2.122\mu m\ H_2$ (blue), $2.26\mu m$ continuum (green) and 
$\rm 2.165\mu m\ Br\gamma $
(red) images is shown in Fig.~\ref{Fig11}.
The emission structures in $\rm Br\gamma$ are clearly visible (almost pure red)
on this composite image and they coincide with the clearly visible (red again)
infrared excess structures on the true color $KHJ$ image (Fig.~\ref{Fig01}).

%

A $\rm Br\gamma$ image with subtracted continuum was generated to outline
better the star forming regions.
The continuum level was determined by measuring the counts for several stars 
common to both the continuum and the line plus continuum images.
The selected from the $JHK$ color-magnitude diagrams stars were main sequence
foreground stars with photometric colors close to those of spectral
class A0.  The calculated ratios of counts were then averaged and the continuum
image was scaled to the line plus continuum image so that the average ratio is
unity. The continuum image was then subtracted from the line plus continuum
image. Since the subtraction of continuum is a very important step we checked our
results using the method of Golev et al. (1996). The resulting emission image obtained 
by this method is practically the same as previous one.  

\begin{figure}[htbp]
	   \resizebox{\hsize}{!}{\includegraphics{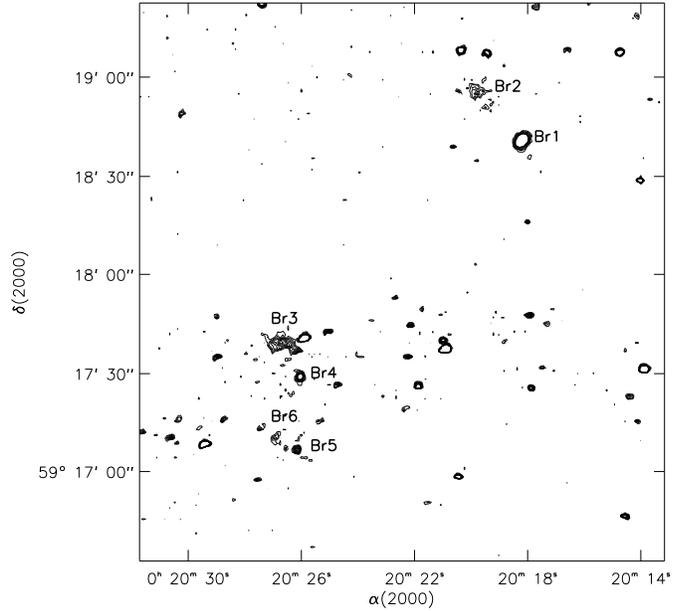}}
 \caption{ Contour plot of continuum-subtracted ${\rm Br\gamma}$ image.
The lowest contour corresponds to 3\,$\sigma$ (2.84 $10^{-16}$ $\rm erg\,sec^{-1}\,cm^{-2}\,arcsec^{-2}$ ), 
the highest contour to 10\,$\sigma$. }
\label{Fig12}
\end{figure}

The emission structures are marked in Fig.~\ref{Fig12}.
The lowest and the highest contours correspond to 3$\sigma$ and 10$\sigma$
respectively. The contour unit is 9.48 $10^{-17}$ $\rm erg\,sec^{-1}\,cm^{-2}\,arcsec^{-2}$ (1$\sigma$
level).
Six "knots" and "spots" with diameters twice greater than FWHM can be located
above the 3$\sigma$ sky value.
Their measured positions in $\alpha_{2000}$, $\delta_{2000}$, 
size in arcsec and fluxes per area per sec are given in Table~\ref{Tab01}.
The  $\rm Br\gamma$ emission sources are rather compact with sizes not larger
than 20 arcsec. Br1, Br2 and Br3 have diffuse morphology while Br4, Br5 and
Br6 are stellar-like objects. The distance between the nearest emission regions
is much larger than the seeing during the observations (1-1.2 arcsec).

%

\begin{table}[h]\tabcolsep=5pt
\caption {Bracket gamma emission structures }
\begin{tabular} {lllrl}
\hline
& & & & \\[-7pt]
Name& $\alpha_{2000}$	 &  $\delta_{2000}$  & \multicolumn{1}{l}{Size} & Flux \\
&$00^{\rm h} + $  & $59^{\circ} + $ & \multicolumn{1}{l}{(arcsec)} &($\rm ergñ\  cm^{-2}\  sec^{-1}$) \\
& & & & \\[-7pt]
\hline
& & & & \\ [-7pt]
Br1 &  $20^{\rm m}\ 18\fs 15$ & $18'\ 39\farcs 9$ & 15.3 & 4.44 $10^{-14}$ \\
Br2 &  $20^{\rm m}\ 19\fs 74$ & $18'\ 55\farcs 2$ & 15.3 & 2.41 $10^{-14}$ \\
Br3 &  $20^{\rm m}\ 26\fs 69$ & $17'\ 37\farcs 6$ & 22.1 & 5.63 $10^{-14}$ \\
Br4 &  $20^{\rm m}\ 26\fs 04$ & $17'\ 28\farcs 4$ & 8.5   & 1.24 $10^{-14}$ \\
Br5 &  $20^{\rm m}\ 26\fs 13$ & $17'\ 05\farcs 4$ & 3.4   & 4.25 $10^{-15}$ \\
Br6 &  $20^{\rm m}\ 26\fs 87$ & $17'\ 09\farcs 6$ & 4.2   & 3.25 $10^{-15}$ \\
& & & & \\[-7pt]
\hline
\end{tabular}
\label{Tab01}
\end{table}

Using the same procedure  as in the $\rm Br\gamma$ reduction, continuum subtracted 
 $\rm H_2$ image was also obtained. There is no detected $\rm H_2$ emissions down to 4.65
$10^{-16}$ $\rm erg\,sec^{-1}\,cm^{-2}\,arcsec^{-2}$  which corresponds to 3\,$\sigma$ level). 
According to Puxley et al. (1995)  if the observed $\rm H_2$/$\rm Br\gamma$ ratios are smaller 
than 0.9 the dominant excitation mechanism of the hot molecular gas is from UV excitation 
by young stars, while if the ratio is larger than 0.9 there is mainly shock excitation.

There is an excellent spatial correlation between the $\rm H\alpha$ (Hodge \& Lee 1990) and the 
$\rm Br\gamma$ emission structures. All $\rm Br\gamma$ emission structures are cross-identified
and in Table~\ref{Tab02} their number and fluxes per area per sec as measured by Hodge \& Lee (1990) 
(see their Table~\ref{Tab02}) are given.  The  $\rm H\alpha$ fluxes are recalculated to the same
size as $\rm Br\gamma$. The comparison between  $\rm Br\gamma$ and  $\rm H\alpha$ fluxes allow us 
to calculate the extinction for each of them using an intrinsic ratio
 $\rm H\alpha$ / $\rm Br\gamma$ = 103.6 (Osterbrock 1989). 
Then using the reddening relation $A_{\rm H\alpha} = 2.32\,E(B-V)$ we calculate color excess. 
The results are given in Column 4 and 5 of Table~\ref{Tab02}.
Taking into account errors in measured fluxes (both in  $\rm Br\gamma$ and  $\rm H\alpha$)
we estimated total error of $10-15\%$.  The mean extinction is $E(B-V)_ {\rm H\alpha}= 1.8\pm0.2$

Since the $\rm H\alpha$ emission is due to ionization from young stars, 
the $\rm H\alpha$ flux is proportional to the star formation rate. Using equation 8 (Miller 1996)
for normal disk and dwarf galaxies with Salpeter IMF the local star formation rates were calculated.
Each of six emission structures is dereddened with  their individual $A_{\rm H\alpha}$.
The SFRs are given in Column 6 of Table~\ref{Tab02}.  
We can also determine the SFR from the  $\rm Br\gamma$ luminosities. In this case we used the calibration equation 9 given in 
Calzetti (1997). To correct the $\rm Br\gamma$ fluxes for the extinction we assume  $A_{K}=0.1\, A_{V}$.  The calculated SFR 
(for star masses 0.1 - 100) are listen in Column  7 of Table~\ref{Tab02}. As can be seen we have very good agreement between SFRs obtained
from  $\rm H\alpha$ and $\rm Br\gamma$ fluxes. 

\begin{table}[h]\tabcolsep=1.5pt\scriptsize
\caption {Bracket gamma -  $\rm H\alpha$ comparison}
\begin{tabular} {llrcccc}
\hline
& & & &&& \\[-7pt]
Name& Hodge & \multicolumn{1}{c}{$\rm H\alpha$ Flux} & $A_{\rm H\alpha}$ &$E(B-V)$& $\rm SFR_{ {\rm H\alpha}}$ &$\rm SFR_{\rm Br\gamma} $\\
& number & $\rm erg\,cm^{-2}\,sec^{-1}$  &  & & $M_\odot\,{\rm year}^{-1}$& $M_\odot\,{\rm year}^{-1}$\\
\hline
& & & &&& \\ [-7pt]
Br1 & E45      & 12.30 $10^{-14}$   & 4.21   &  1.81  & 0.222 &0.179 \\
Br2 & E50      & 7.94 $10^{-14}$   & 3.98   &  1.72  & 0.116 &0.095 \\
Br3 & E111c,d & 11.49 $10^{-14}$ & 4.21  &  1.82  & 0.207 &0.227\\
Br4 & E111a   & 1.74 $10^{-14}$   & 4.37   &  1.88  & 0.036 &0.051\\
Br5 & E106a   & 0.40 $10^{-14}$   & 5.05   &  2.18  & 0.016 &0.018\\
Br6 & E106b   & 1.78 $10^{-14}$   &  3.36   & 1.45  & 0.015 &0.012\\
& & & &&& \\[-7pt]
\hline
\end{tabular}
\label{Tab02}
\end{table}

The summarized SFR derived from six $\rm H\alpha$ and  $\rm Br\gamma$ emission structures in our field of view is 
$0.61 M_\odot {\rm year^{-1}}$ and $0.58 M_\odot {\rm year^{-1}}$ respectively.
 SFR for the whole galaxy given in Table\,5 of Mateo (1998) is $0.71 M_\odot {\rm year^{-1}}$. 
Our field covers only the active area of IC\,10 and hence the SFR of the whole galaxy must be higher, which confirms its starburst status.

\section{Summary}

The infrared photometry presented here together with the $\rm Br\gamma$
recombination line images provide useful information on the star-formation
episodes of the Local group dwarf irregular galaxy IC\,10.  

The analysis of $(J-K, K)$
and $(H-K,K)$ color-magnitude diagrams clearly shows the presence of red supergiants,
asymptotic giant branch stars and red giant branch stars.
The red supergiants have masses between 15 and 30 $M_\odot$
and ages between 10 and 50  Myr. We would like to point out the presence of
a significant amount of  young (ages less than 1 Gyr) AGB star population.

Comparing the red supergiants of IC\,10 and IC\,1613
we have determined the reddening of $E(B-V)=1.05\pm0.10$ 
and the dereddened distance modulus $(m-M)_{0}=23.86\pm0.12$ mag of
 Population I stars in IC\,10.

The detection of six  $\rm Br\gamma$  emission structures clearly outlines
two star forming regions in our field of view.
There is no $\rm H_2$ emission down to 4.65 
$10^{-16}$ $\rm erg\,sec^{-1}\,cm^{-2}\,arcsec^{-2}$ indicating that the dominant excitation mechanism of the  molecular gas comes from UV radiation
from hot young stars.

From the observed  $\rm Br\gamma$ and
$\rm H\alpha$ fluxes, we derive the average extinction toward the star forming regions
$E(B-V)_ {\rm H\alpha}= 1.8\pm0.2$. 
This extinction refers to the ionized gas and is larger than the stellar reddening 
derived from the red supergiants. The  $\rm Br\gamma$ sources are likely to be those with the highest
surface brightness.  As a consequence, they will tend to be compact
H II regions, and compact H II regions are embedded in
molecular clouds, usually deeply embedded.  As a result, they are
also highly reddened, which would explain why the nebular reddening
is  higher than stellar reddening.

 The high SFR derived from six $\rm H\alpha$ and  $\rm Br\gamma$ emission structures in our field of view confirms starburst status of IC\,10.

\begin{acknowledgements}
L. G. and J. B. would like to thank Anabel Arieta and Alicia Porras
for their help in the process of reducing the IR frames and also
E. Chelebiev for his help. The authors gratefully acknowledge the useful
comments and suggestions raised by Drs. F. Ferraro, M. R. Cioni, M. Richer,
V. Golev, T. Bonev and V.D. Ivanov as well as the comments by an anonymous 
referee.
This work was performed while J.B. was a
visiting astronomer in UNAM, Mexico under contacts CONACYT No.400354-5-2398PE
and DGAPA INI04696 and was supported
in part by the Bulgarian National Science
Foundation grant under contract No. F-812/1998 with
the Bulgarian Ministry of Education and Sciences.
\end{acknowledgements}

\newpage

 \end{document}